\documentstyle[psfig]{caosp}
\def\LB{$\lambda$\,Bootis }

\begin{document}
\hauthor{E. Paunzen}
\title{On the evolutionary status of \LB stars using Hipparcos data}
\author{E. Paunzen} 
\institute{Institut f\"ur Astronomie der Universit\"at Wien, T\"urkenschanzstr.
17, A-1180 Wien (paunzen@galileo.ast.univie.ac.at)}
\maketitle
\begin{abstract}
Using the Hipparcos data, absolute magnitudes and thus
the evolutionary status for the group of \LB stars were
derived. The origin of this small
group of non-magnetic, chemically peculiar stars, still remains a matter of
debate. Using new evolutionary tracks, an age determination could be provided
to distinguish between the two competing theories - the
diffusion/mass-loss and the accretion.

The results establish the members of this group as objects which are {\it
very close to the Main Sequence}. This is supported by Pre-Main Sequence
evolutionary tracks as well as by observational results. This
contradicts prior conclusions that most of these stars are in the middle of
their Main Sequence lifetime. The new results strongly support the
predictions of the accretion theory.

\end{abstract}

\section{Introduction}

Using the Hipparcos parallaxes, absolute
magnitudes for members of the \LB group were 
determined.
The origin of these non-magnetic,
A to F-type metal-deficient dwarfs still remains controversial. The two main
competing theories involve diffusion, either
in combination with mass-loss (Michaud \& Charland 1986), or
accretion of interstellar matter as in post-AGB stars (Turcotte \&
Charbonneau 1993). The latter model requires that \LB stars are very
close to the Zero-Age Main Sequence. The recent discovery of \LB stars
in the young Orion OB1 association and in NGC~2264 (Paunzen \& Gray 1997)
seems to support the predictions of the accretion theory. But an age
determination of
galactic field stars by Iliev \& Barzova (1995, hereafter IB95), on the other 
hand, resulted in evolved members. These stars were recalibrated using the
accurate new absolute magnitudes as well as new stellar evolutionary
tracks (CESAM; Morel 1997). Furthermore, Pre-Main Sequence
models (Palla \& Stahler 1993) were used to confirm the results.

\section{The new Hipparcos data} \label{evol}

Candidates as well as members of the \LB group were taken from Paunzen
et al. (1997) and Paunzen \& Gray (1997). The Hipparcos data for the
programme stars were extracted with the help of Simbad.
The observed visual magnitudes were used to calculate the absolute magnitudes
$M_{\rm V}$(H).

Possible correlations of the observed parallaxes with other astrophysical
quantities (e.g. apparent distance, effective temperature, metallicity, etc.)
were examined (see also Paunzen 1997).
No systematic trend between the (old) photometrically
calibrated and (new) absolute magnitudes has been detected. Although there
are some individual differences, the overall validity of the ``standard''
calibration for the (chemically peculiar) \LB stars is proven.

IB95 presented an age and mass determination for 20 well
established \LB stars (and Vega). They concluded that most of the investigated
stars are in the middle of their Main Sequence evolution, which is believed to
be inconsistent with the much favoured accretion theory. Only one star of
their sample (HD~290799, a member of the young Orion OB1 association) seems to
fulfill the predictions of the accretion theory. The
Hipparcos data (available for 18 stars from IB95) were
used to
test their conclusions. 

After the calibration of the programme stars in a $log T_{\rm eff}$ versus
$log L/L_{\sun}$ diagram, the new CESAM models 
(Morel 1997) were used to determine the ages and masses.
The initial parameters of the evolutionary tracks were
X\,=\,0.7 and Z\,=\,0.02 (solar abundance); these values were found to be
valid for the study of (chemically peculiar) \LB stars by IB95.
This seems to be appropriate because the main contribution to the
overall metallicity is due to C, N and O (solar abundant in \LB stars).
Furthermore, there are strong indications that the \LB phenomenon is
restricted to the stellar surface (Holweger \& Rentzsch-Holm 1995).

Due to the individual corrections to the
absolute magnitude, all programme stars (except HD~193256 and HD~193281,
a distant close binary system) are, within the errors, significantly
{\it younger} and {\it less massive}. In order to test a possible
Pre-Main Sequence hypothesis and thus the consistency with the accretion
theory, the evolutionary tracks from Palla \& Stahler (1993) were
applied.

It turned out that six stars (HD~30422,
HD~31295, HD~107233, HD~110411, HD~125162 (\LB itself) and HD~183324) are
indeed {\it very
close to the Main Sequence}. This is proven by the
individual results derived from the Pre- and Main Sequence tracks.
For three additional stars (HD~38545, HD~111786 and
HD~221756), both models are very close, resulting in the
same conclusion with a high confidence. These findings contradict the results 
from IB95 and strongly support the accretion theory.
The remaining nine programme stars are on the Main Sequence (using the 
appropriate models) but still significantly less evolved than reported by
IB95 (see Fig. 2 therein).

These results are consistent with values for
other low-mass ``dusty'' Pre-Main Sequence objects (see also
Gerbaldi \& Faraggiana 1993) such as Herbig Ae/Be stars or $\beta$ Pictoris.
The explanation for \LB stars as true Pre-Main Sequence object could also lead
to a solution for the apparent small number of members. A star with
2\,M$_{\sun}$ needs only a few 10$^{6}$ years to reach the Main Sequence.
The probability to find such objects is therefore very small compared to the
lifetime on the Main Sequence (there is also only a similarly small number
of Herbig Ae stars known). This conclusion is further strengthened by the lack
of \LB stars in open clusters older than 10$^{7}$ years (Paunzen \& Gray 1997).

\section{Conclusions}

With the Hipparcos data, absolute magnitudes and
evolutionary status (mass and age) for the group of \LB stars were
estimated. No systematic influence of the apparent distance, effective 
temperature, metallicity and rotational velocity was found on the difference 
between the photometrically calibrated and the ``new'' absolute magnitudes,
thus proving the validity of the ``standard'' photometric calibration
(e.g. in the Str\"omgren system) for these (chemically peculiar) stars.

It turned out that six stars
(e.g. \LB itself) are definitely {\it very close to the Main Sequence},
and this is also true, with a high probability, of three additional programme 
stars. These results contradict the conclusions of IB95
and support the accretion theory. The small number of \LB stars (statistical
effect due to the short ``lifetime'' on the Pre-Main Sequence) and the lack
of them in open clusters older than 10$^{7}$ years further strengthen
the accretion hypothesis.

\acknowledgements
This research was carried out\,within\,the working group
{\it Astero-} {\it seismology-AMS } with
funding from the Fonds zur F\"orderung der wissenschaftlichen
Forschung (project {\em S7303-AST}).
Use was made of the Simbad database,
operated at CDS, Strasbourg, France.

\end{document}